\documentclass[12pt]{iopart}
\usepackage{epsfig}
\usepackage{cite}
\usepackage{amsfonts}
\usepackage{times}
\begin{document}
\title[Specific heat of MgB$_2$ from first-principles calculations]
{Specific heat of MgB$_2$ in a one- and a two-band model
from first-principles calculations}

%
%
\author{A A Golubov\dag,  J Kortus\ddag,
O V Dolgov\ddag, O Jepsen\ddag, 
Y Kong\ddag, O K Andersen\ddag,
B J Gibson\ddag, K Ahn\ddag\ and R K Kremer\ddag
}
%
%
\address{
  \dag University of Twente, Department of Applied Physics,
   7500 AE Enschede, The Netherlands\\
  \ddag Max-Planck-Institut f\"ur Festk\"orperforschung,
   Heisenbergstr. 1, D-70569 Stuttgart, Germany\\
}
\date{\today}
\ead{j.kortus@fkf.mpg.de}
%
%
\begin{abstract}
The heat capacity anomaly at the transition to superconductivity
of the layered superconductor MgB$_2$ is
compared to first-principles calculations
with the Coulomb repulsion, $\mu^\ast$, as the only parameter which
is fixed to give the measured $T_c$. 
We solve the Eliashberg equations for both an isotropic one-band and 
a two-band model with different superconducting gaps on the 
$\pi$-band and $\sigma$-band Fermi surfaces. The agreement with
experiments is considerably better for the two-band model than for the
one-band model.
\end{abstract} 
\submitto{\JPCM}
\pacs{74.25B, 74.80Dm, 74.72.-h}
\maketitle

\section{Introduction}

The nature of the superconducting state in MgB$_2$ has been characterized 
by a broad range of experimental and theoretical methods and 
many basic properties have been  unambiguously established  
since the discovery of the 40\,K superconductor MgB$_2$ by
Nagamatsu and collaborators \cite{Akimitsu}. 

While  electron-phonon coupling as the underlying pairing mechanism
has been pinpointed by a large B isotope effect on  $T_c$
proving B related vibrations to be essential \cite{Budko,Hinks}
the nature of the order-parameter (viz.\ the superconducting gap)
has remained a matter of debate.  The order-parameter has been
intensively investigated by tunnelling and point contact
spectroscopy
\cite{Karapetrov,Schmidt,Rubio,Sharoni,Kohen,Plecenik,%
Szabo,Giubileo1,Giubileo2,Chen,Laube,Zhang,Bugoslavsky} as well as
by high-resolution photoelectron spectroscopy
\cite{Takahashi,Tsuda}. While these techniques show  an energy gap
in the quasiparticle spectrum most likely of $s$-wave symmetry the
magnitude of the gap, $\Delta(0)$, itself remained an open
question: tunnelling experiments initially revealed a distribution
of energy gaps with lower boundary 2$\Delta_1(0)$/$k_{\rm B}T_c
\approx$\,1.1 and upper boundary 2$\Delta_2(0)$/$k_{\rm B}T_c
\approx$\,4.5. 
These values are either considerably lower or
distinctly larger than the weak coupling BCS value of
2$\Delta(0)$/$k_{\rm B}T_c=$\,3.53 and these
controversial findings have been discussed in terms of gap
anisotropy or more recently attributed to the presence of two
gaps or multiple gaps \cite{Giubileo2,Tsuda}. The analysis of the
electronic Raman continuum of MgB$_2$ by Chen {\it et al.} \cite{Chen} 
also pointed to the presence of two gaps with gap values within the
limits indicated by the tunnelling experiments \cite{Franck}.

While these experiments employ surface sensitive
techniques to determine the gap properties, evidence for multigap
behaviour emerges also from methods like heat capacity or $\mu SR$
measurements probing true bulk properties \cite{classical}. In the
early heat capacity experiments the typical jump-like anomaly is
seen at $T_c$ the magnitude $\Delta C_p$/$T_c$ of
which amounts at best to only about 70-80\,\% of the value
1.43\,$\gamma^{\rm N}(T_c)$ predicted by weak-coupling BCS
theory \cite{Budko,Kremer,Bouquet1,Ott,Wang,Yang}. 
$\gamma^{\rm N}(T_c)$ is the Sommerfeld constant in the normal state which was
obtained from heat capacity measurements in high magnetic fields
and which was determined to be 2.7-3 mJ mol$^{-1}$\,K$^{-2}$. The shape of the 
heat capacity anomaly compares reasonably well with BCS-type behaviour
assuming 2$\Delta(0)$/$k_{\rm B}T_c$\,=\,3.53 with
appropriately adjusted magnitude. An improved fit of the detailed
temperature dependence of the heat capacity anomaly was obtained
when calculating the heat capacity within the
$\alpha$-model \cite{Padamsee} assuming a BCS temperature
dependence of the gap but with an increased ratio
2$\Delta(0)$/$k_{\rm B}T_c$\,=\,4.2(2) \cite{Kremer}. This
result matches very well with the upper limit of the gap value
consistently found in the tunnelling experiments and was suggested
as an evidence that MgB$_2$ is in the moderately strong coupling
limit. More recently, the excess heat capacity observed close to
$T_c$/4 by Bouquet {\it et al.} \cite{Bouquet1} and Wang \etal.
\cite{Wang} 
has been attributed to a second smaller gap. Fits with a
phenomenological two-gap model assuming that the heat capacity of
MgB$_2$ can be composed as a sum of the two individual heat
capacities gave very good description with gap values of
2$\Delta_1(0)$/$k_{\rm B}T_c$\,=\,1.2(1) and
2$\Delta_2(0)$/$k_{\rm B}T_c\,\approx$ 4
\cite{Bouquet2}.
Recent muon-spin-relaxation measurements of the magnetic
penetration depth are consistent with a two-gap model
\cite{Niedermayer}.

Theoretically multigap superconductivity in MgB$_2$ was first 
proposed by Shulga {\it et al.} to explain  the behaviour of the
upper critical magnetic field \cite{Shulga}. Based on the
electronic structure the existence of multiple gaps has been
suggested by Liu {\it et al.} in order to  explain the magnitude
of $T_c$ \cite{Liu}. The electronic structure of MgB$_2$
contains four Fermi surface sheets \cite{Kortus}. Two of them with
2D character emerging from bonding $\sigma$ bands form small
cylindrical Fermi surfaces around $\Gamma$\,-\,A. The other two
originating from bonding and antibonding $\pi$ bands have 3D
character and form a tubular network. Liu {\it et al.} from
first-principles calculations of the electron-phonon coupling
conclude that the superconducting gap  is different for the
individual sheets and they obtain two different order parameters,
a larger one on the 2D Fermi surface sheets and a second gap on
the 3D Fermi surfaces, the latter was estimated to be approximately
a factor of three reduced compared to the former \cite{Liu}.

In the present paper we calculate the specific heat capacity from the
spectral Eliashberg function $\alpha^2(\omega)F(\omega)$ first
in the one-band model using the isotropic $\alpha^2(\omega)F(\omega)$ 
as given by Kong {\it et al.}\cite{Kong}. Then we calculate the heat
capacity in a two-band model by reducing the 16 Eliashberg
functions $\alpha^2_{ij}(\omega)F_{ij}(\omega)$ appropriate for the
four Fermi surface sheets into four Eliashberg functions
corresponding to an effective-two-band model with a $\sigma$-band and
$\pi$-band only. From the solution of the Eliashberg equations
we obtain a superconducting gap ratio 
$\Delta_{\sigma}/\Delta_{\pi}\,\simeq 2.63$\,
in good agreement with the experimental data. 
The two-band model explains the reduced magnitude of the
heat capacity anomaly at $T_c$ very well and also  reproduces the
experimental observed excess heat capacity at low temperatures.

\section{Theory}
\subsection{One-Band Model}
First we discuss the specific heat in the isotropic single
band model with a strong (intermediate) electron-phonon
interaction (EPI). In the normal state and in the adiabatic
approximation the electronic contribution to the specific heat is
determined from the Eliashberg function $\alpha ^{2}(\omega
)F(\omega )$ by the expression \cite{Grimvall}
\begin{eqnarray}
C_{N}^{el}(T) &=&(2/3)\pi ^{2}N(0)k_{B}^{2}T  \label{cn} \\
&&\times \left[ 1+(6/\pi k_{B}T)\int_{0}^{\infty }f(\omega /2\pi
k_{B}T)\alpha ^{2}(\omega )F(\omega)\rmd\omega \right] ,  \nonumber
\end{eqnarray}
where $N(0)$ is a bare density of states per spin at the Fermi
energy. The kernel $f(x)$ is expressed in terms of the derivatives
of the digamma function $\psi (x)$
\begin{equation}
f(x)=-x-2x^{2} \mathop{\rm Im} \psi ^{\prime }(\rmi x)-x^{3}
\mathop{\rm Re}\psi^{\prime\prime }(\rmi x). 
\label{func}
\end{equation}

At low temperatures the specific heat
has the well known asymptotic form: $C_{N}^{el}( T\rightarrow
0)=(1+\lambda )\gamma _{0}T$, where $\lambda =2\int_{0}^{\infty
}d\omega\,\omega ^{-1}\alpha
^{2}(\omega )F(\omega )$ is the electron-phonon coupling constant, and $%
\gamma _{0}=2\pi ^{2}k_{B}^{2}N(0)/3$ is the specific heat
coefficient for noninteracting electrons. At higher temperatures
the specific heat differs from this trivial expression  
(see, the discussion in reference \cite{sdm}).

In the superconducting state an expression for the specific heat
obtained by Bardeen and Stephen \cite{Bardeen} which is based on
an {\it approximate} sum rule has often been used.
We shall however use an {\it exact} expression for the
thermodynamical potential in the electron-phonon system which is
based on the integration of the electronic Green's function over
the coupling constant 

\begin{equation}
 \Omega =\Omega _{\rm{el}}^{(0)}+\Omega
_{\rm ph}^{(0)}+T\sum_{\omega _{\rm
n}}\int\limits_{0}^{1}\frac{\rmd x}{x}\,{\rm tr}\left[
\hat\Sigma (x)\hat G(x)\right]
\end{equation}
where $x$ is dimensionless, 
$\hat{G}(x)$ and $\hat \Sigma (x)$
are the exact electron Green's function and the self-energy,
respectively, for a coupling constant of  $x\cdot \lambda $.
The functions
$\Omega _{\rm el}^{(0)}$ and $\Omega _{\rm ph}^{(0)}$  are the
thermodynamic potentials for noninteracting electrons and
noninteracting phonons, respectively. Some further arithmetics
leads to the expression for the difference in free energies,
$F_{\rm N}$ and $F_{\rm S}$, of the superconducting and normal
state \cite{gd}

\begin{equation}
-\frac{F_{N}-F_{S}}{\pi N(0)T}=
\sum\limits_{n=-\omega _{c}}^{\omega
_{c}}\left\{
\begin{array}{c}
|\omega _{n}|(Z^{N}(\omega _{n})-1)-\frac{2\omega _{n}^{2}[\left(
Z^{S}(\omega _{n})\right) ^{2}-1]+\varphi _{n}^{2}}{|\omega _{n}|+\sqrt{%
\omega _{n}^{2}\left( Z^{S}(\omega _{n})\right) ^{2}+\varphi
_{n}^{2}}} \\ +\frac{\omega _{n}^{2}Z^{S}(\omega
_{n})(Z^{S}(\omega _{n})-1)+\varphi _{n}^{2}}{\sqrt{\omega
_{n}^{2}\left( Z^{S}(\omega _{n})\right) ^{2}+\varphi _{n}^{2}}}
\end{array}
\right\} ,  \label{free}
\end{equation}
where $Z(\omega_n)$ is a normalization factor,
$\varphi_{n}=\Delta_{n}/Z(\omega_n)$ is an order parameter,
and $\Delta_{n}$ is the gap function.

The specific heat at temperature, $T$, is calculated according to:
\begin{equation}
\Delta C_{\rm el}(T)=T\partial ^{2}(F_{N}-F_{S})/\partial T^{2}.
\label{partial}
\end{equation}
The specific heat jump $\Delta C_{\rm el}(T_c)$ at $T=T_c$ 
is determined by the coefficient $\beta
=T_{c}\Delta C_{\rm el}(T_{c})/2$ in $F_{N}-F_{S}=\beta t^{2}$, where 
$t=(T_{c}-T)/T_{c}$.

\subsection{Two-Band Model}
We have calculated  the 16  Eliashberg functions
$\alpha _{ij}^{2}(\omega )F_{ij}(\omega )$ where $i$ and $j$ label
the four Fermi surface sheets and thereafter combined them into
four corresponding to an {\it effective} two-band model
which contains only a $\sigma$- and $\pi$-band. 
Their respective
densities of states at the Fermi energy have values  
of $N_{\sigma}(0)=0.300$ states/cell$\cdot $eV and 
$N_{\pi}(0)=0.410$ states/cell$\cdot $eV. 
Similar coupling constants $\lambda_{\sigma\sigma},
\lambda_{\pi\pi}, \lambda_{\sigma\pi}$, and 
$\lambda_{\pi\sigma}$ which are required for a two-band model
were calculated earlier in reference\cite{Liu}. 
The procedure of reducing the 16 Eliashberg functions of the
real 4 band system due to the 4 different Fermi surface sheets
to an effective two-band model with only four coupling constants
$\lambda_{ij}$ is an approximation which is based on the similarity
of the two cylindrical and the two three-dimensional sheets of the
Fermi surface requiring the same physical properties in both
$\sigma$-bands or both $\pi$-bands. More details can be found 
elsewhere\cite{Dolgov-optic}.

\begin{figure}
\centerline{\psfig{figure=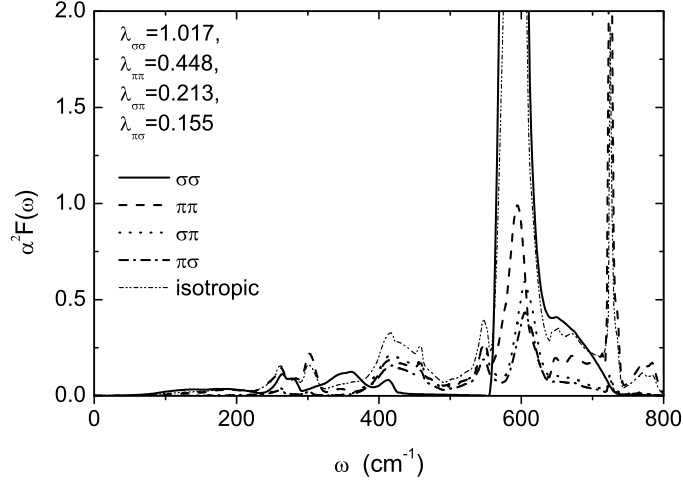,width=10cm,angle=0}}
\caption{The four superconducting Eliashberg functions 
$\alpha^2\,F(\omega)$ obtained from first-principles calculations
for the effective two-band model and the
isotropic Eliashberg function for the one-band model.
The coupling constant of the isotropic one-band model has a value of
$\lambda_{iso}$=0.87.
}
 \label{fig:eliash}
\end{figure}

The four Eliashberg functions $\alpha _{ij}^{2}(\omega )F_{ij}(\omega)$
for the effective two-band model are shown in figure~\ref{fig:eliash}. 
The most significant contribution comes from the coupling of the
bond stretching phonon modes to the $\sigma$-band. The 
coupling constants corresponding to the superconducting
Eliashberg functions have been calculated to be: 
$\lambda_{\sigma\sigma}=1.017,
\lambda_{\pi\pi}=0.448, \lambda_{\sigma\pi}=0.213,$ and 
$\lambda_{\pi\sigma }=0.155$. 
The small difference to the values given in reference \cite{Liu} 
may be attributed to the different first-principles methods 
used in the calculation of the Eliashberg functions. 

Besides the spectral functions we need to know the
the Coulomb matrix element $\mu_{ij}$. With the help of the wavefunctions
from our first-principles calculations we can approximately calculate
the ratios for the $\mu$-matrix \cite{Dolgov-optic}.
The $\sigma\sigma$-, $\pi\pi$- and $\sigma\pi$-values were in the ratio 
2.23/2.48/1.
This allows one to express $\mu^\ast_{ij}(\omega_c)$ by these ratios
and one single free parameter which is fixed to get 
the experimental $T_c$ of 39.4 K from the solution of the Eliashberg
equations. The $\mu^\ast(\omega_c)$ matrix elements determined by this
procedure are $\mu^\ast_{\sigma\sigma}(\omega_c)$=0.210,
$\mu^\ast_{\sigma\pi}(\omega_c)$=0.095, 
$\mu^\ast_{\pi\sigma}(\omega_c)$=0.069, and
$\mu^\ast_{\pi\pi}(\omega_c)$=0.172.

\begin{figure}
\centerline{\psfig{figure=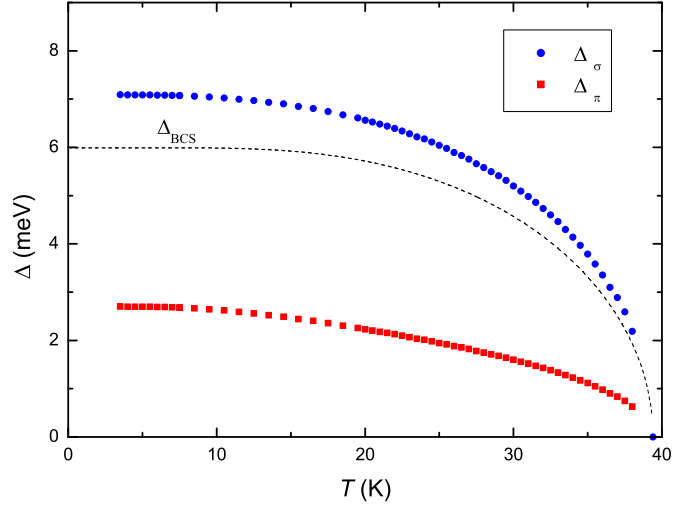,width=10cm,angle=0}}
\caption{The temperature dependence of the the superconducting
gaps from the solution of the two-band Eliashberg equations.
The values of the gaps at $T$=0 K were obtained as  
$\Delta_\sigma \left(T=0\right)$ = 7.1 meV and 
$\Delta_\pi \left(T=0\right)$ = 2.7 meV. The BCS value for the gap
that corresponds to $T_c$=39.4 K is 6.0 meV at 0 K.
}
\label{delta2}
\end{figure}

Using our calculated Eliashberg functions 
on the imaginary (Matsubara) axis together with the above 
matrix $\mu^{\ast}_{ij}(\omega_c)$ we obtain the gap values 
$\Delta_{\sigma}=\lim\limits_{T\to 0}\Delta_\sigma(i\pi T)\simeq 7.1\,$meV,
and $\Delta_{\pi}\simeq 2.7\,$meV, 
which corresponds to $2\Delta_\sigma/k_B T_c$=4.18 and
$2\Delta_\pi/k_B T_c$=1.59.
The temperature dependence of the superconducting gaps is shown
in figure~\ref{delta2}. The filled circles (squares) display the
gap for the 2D $\sigma$- (3D $\pi$-) band.
Due to interband coupling between the bands both gaps close at the same 
critical temperature.
For a comparison also the BCS curve (line) is shown
for a single gap (one-band model) which closes at $T_c$=39.4 K.
The corresponding single BCS gap would be 6 meV. 

The extension of equation \ref{cn} to
the two-band model gives
\begin{equation}
C_{\rm el}^{\rm N}(T)=\frac{2\pi ^{2}}{3}N_{tot}(0)k_B^{2}T+
{4\pi}{k_B}
[N_{\sigma }(0)(I_{\sigma\sigma }+I_{\sigma\pi})+
N_{\pi}(0)(I_{\pi\pi}+I_{\pi\sigma})]
 \label{twoband}
\end{equation}
where $I_{ij}=\int_0^\infty f(\omega/2\pi k_BT)
\alpha_{ij}^2(\omega)F_{ij}(\omega)d\omega$ 
($i,j=\pi ,\sigma$), and the function $f(x)$ is given by
equation \ref{func}.

The generalization of the superconducting free energy (\ref{free})
to the two band model is straightforward and the heat capacity was
obtained according to equation \ref{partial}.

\section{Comparison with Experiment}
\begin{figure}
\centerline{\psfig{figure=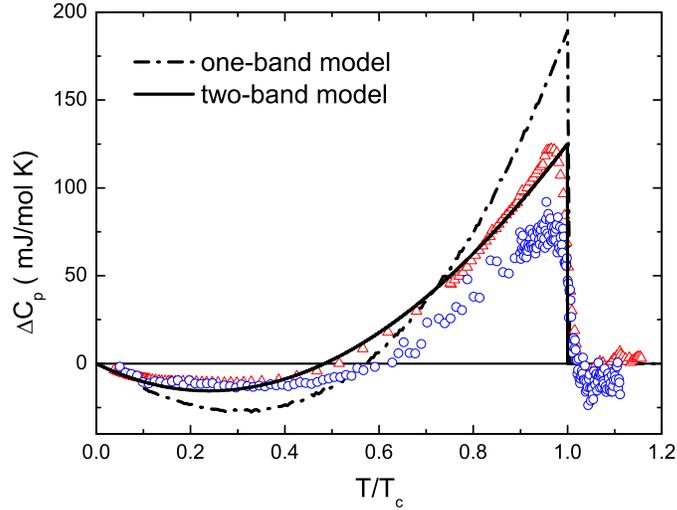,width=10cm,angle=0}} 
\caption{Experimental data of the
heat capacity difference $\Delta C_p=C_p(0{\rm Tesla})-C_p(9{\rm Tesla})$ 
 from reference~\protect\cite{Kremer} ($\circ$) and
from reference~\protect\cite{Bouquet1} ($\opentriangle$). 
The dashed line is the theoretical result of the one-band model and the 
thick solid line corresponds to the two-band model from the solution
of the Eliashberg equations.
The two-band model reproduces much better the specific heat jump as well
as the low temperature behaviour.
}
\label{fig:cp}
\end{figure}

For the comparison with experiment we have selected data
obtained by our group \cite{Kremer} and by 
Bouquet \etal.~\cite{Bouquet1}.
The anomaly clearly visible at $T_c$ in the zero-field data
is suppressed by a magnetic field of 9 Tesla in both experiments.
In figure \ref{fig:cp} we display the difference
$\Delta C_p=C_p(0{\rm Tesla})-C_p(9{\rm Tesla})$. 
The anomalies at $T_c$ detected by both
groups clearly have a different magnitude, the one described in reference 
 \cite{Bouquet1} amounts to 133 mJ/mol K at $T_c$ and represents
the largest specific heat capacity anomaly reported for MgB$_2$ 
so far \cite{exp}.
The $\Delta C_p(T_c)$ reported by our group is somewhat smaller,
however, the shape of the anomalies close to $T_c$ is very similar 
for both samples. In fact, fitting the anomalies with the $\alpha$-model 
revealed an identical ratio $2\Delta/k_BT_c$=4.2 with $\Delta$=7 meV 
for both samples \cite{Kremer,Bouquet1,Bouquet2}.

First we will try to discuss the experimental results in terms
of a conventional one-band model.
The specific heat in MgB$_{2}$ 
was calculated using the isotropic spectral Eliashberg function 
$\alpha^{2}(\omega)F(\omega)$ of Kong {\it et al.} \cite{Kong}.
This function yields an electron-phonon coupling constant
$\lambda =0.87$ and together with a Coulomb pseudopotential of
$\mu^{\ast}=0.1$ yields $T_{c}=40$ K. 
The calculated specific heat at $T_{c}$ is 
$\gamma^{\rm N} (T_c)=1.94 \gamma_0=3.24$ mJ/mol\,K$^{2}$ with 
$\gamma _{0}$=1.67 mJ/mol\,K$^{2}$
from the band structure calculations of reference~\cite{Kong,Kortus}.
The specific heat jump at $T_c$ equals $\Delta C\simeq$ 196 mJ/mol K, 
which is a factor of 1.5-2 larger than the experimental values \cite{exp}. 
It corresponds to $\Delta C/(\gamma^{\rm N} (T_c)T_c)\simeq $ 1.51
compared to the BCS value of 1.43. 
The difference 
$\Delta C_{\rm el}(T)= C_{\rm el}^{S}(T)- C_{\rm el}^N (T)$ 
is shown in figure~\ref{fig:cp} (dashed-dot line) in comparison with
the experimental data. 
Not only the size of the jump disagrees with the experiment,
but also the behaviour at low temperatures is different. The
latter is connected with the fact that at low temperature equation
(\ref{free}) for a single band model leads to the standard
exponential dependence $C^{S}\sim T^{-3/2}\exp (-\Delta /T)$,
while the experimental data show a more complicated behaviour.
Clearly there exists a discrepancy between experimental data
and a theoretical one-band model.

The solid line in figure\ \ref{fig:cp} represents the theoretical
results for the two-band model as described above. The low
temperature behaviour is in much better agreement with
the experiment. The specific heat jump is now significantly reduced
in comparison with a single band model and reproduces surprisingly
well the experimental data of reference\ \cite{Bouquet1}.
With the data given above we obtain from our theoretical 
calculation  an electronic heat capacity in the normal state of 
$\gamma^{\rm N}(T_{c})=C_{\rm el}^{\rm N}(T_{c})/T_{c}\simeq 
3.24$ mJ/mol K$^2$, the same value as for the one-band model.
The absolute value of the specific heat jump in the two-band
model is $\Delta C \simeq$ 125 mJ/mol K, corresponding to
$\Delta C/(\gamma^{\rm N}(T_c) T_c)\simeq$ 0.98 which is
now smaller than the BCS value.

We would like to emphasize here that no fitting is involved in the
theoretical calculations. The only free parameter which is in
the Coulomb matrix elements is already determined by the experimental
$T_c$ of 39.4 K. 

One could expect that the difference between the theoretical results
of the effective two-band model and our experimental data may be
attributable to a different amount of impurities in the samples
compared to the samples of reference \cite{Bouquet1}.
In the one-band model the critical temperature $T_c$ as well as
the value and the temperature dependence of $\Delta C_p(T)$ are not
affected by non-magnetic impurities (Anderson theorem). 
This is in complete contrast to the situation for
the two-band model, where both quantities 
are strongly dependent on {\em interband} impurity scattering. 
Interband impurity scattering leads to averaging of the gaps 
and thus to the increase of $\Delta C_p/\gamma^{\rm N}(T_c) T_c$ ratio. 
On the other hand, due to decrease of $T_c$, the specific heat jump 
only depends weakly on the scattering strength. 
In order to investigate the dependence of $T_c$ and of $\Delta C_p$
on the interband impurity scattering we included the 
effect of interband impurities in the Eliashberg equations.
The results show that even for rather strong impurity scattering
$1/2\tau=3\pi T_{c0} \simeq 371$ K, which leads to a drastic change of the
critical temperature (decreasing to $T_c$=29.4 K) and strong averaging 
of the gaps, the specific heat jump remains practically unchanged 
$\Delta C_p\simeq$ 120 mJ/mol K. 
This corresponds to a ratio $\Delta C_p/\gamma(T_c) T_c \simeq$ 1.48, 
which is very
close to the corresponding value of a single gap model. 
Therefore, interband impurity scattering can explain the change of
$T_c$ in different samples, but is not responsible for the observed different
values of the specific heat capacity anomaly at $T_c$.

We have shown that a complete theoretical calculation
from first-principles using an effective two-band model
can explain the major features in the specific heat measurement of 
MgB$_2$ surprisingly well.
The presented theoretical framework goes beyond a simple
phenomenological two-gap model because interband effects are included 
explicitely and no fitting to experimental results has been performed. 
The reduced value of the heat capacity anomaly at $T_c$ as well as
the low temperature behaviour are in excellent agreement with
experimental results.
The same first-principles approach using exactly the same
Eliashberg functions and Coulomb matrix elements  has been used in
order to explain optical measurements \cite{Dolgov-optic} and
tunnelling experiments \cite{Golubov} 
of the interesting superconductor MgB$_2$.

\ack
JK would like to thank the Schloe{\ss}mann Foundation for
financial support.


\begin{thebibliography}{99}
\bibitem{exp}  
It is interesting that experiments on different samples show
similar values for the specific heat jump 
$\Delta C\simeq 113$mJ/mol K \ 
\cite{Budko},
$\Delta C\simeq 81$mJ/mol K \ 
\cite{Wang},
$\Delta C\simeq 133$mJ/mol K \ 
\cite{Bouquet1},
$\Delta C\simeq 125 $mJ/mol K 
\cite{Ott},
$\Delta C\simeq 115$mJ/mol K 
\cite{Yang},
$\Delta C\simeq 92$mJ/molK 
\cite{Fred}
\end{thebibliography}
\end{document}